# A refined numerical investigation of a large equivalent shallow-depth underwater explosion

Ming He(贺铭)[1,2], Shuai Zhang(张帅)[2,3]*, Shi-ping Wang(王诗平)[2,3], Ze-yu Jin(金泽宇)[4], Hemant Sagar[5]

1. State Key Laboratory for Turbulence and Complex Systems, Department of Mechanics and Engineering Science, BIC-ESAT, College of Engineering, Peking University, Beijing 100871, China
2. College of Shipbuilding Engineering, Harbin Engineering University, 145 Nantong Street, Harbin 150001, China
3. Nanhai Institute of Harbin Engineering University, Sanya 572024, China
4. Green & Smart River-Sea-Going Ship, Cruise and Yacht Research Center, Wuhan University of Technology, Wuhan 430063, Hubei, China
5. Institute of Ship Technology, Ocean Engineering and Transport Systems, University of Duisburg-Essen, Bismarckstr. 69, 47057, Duisburg, Germany

Authors to whom correspondence should be addressed: zhangshuai8@hrbeu.edu.cn

**Abstract:** The large equivalent shallow-depth explosion problem is very significant in the field of naval architecture and ocean engineering, as such explosions can be used to attack and demolish ships and anti-ship missiles. In the current work, a refined numerical study of the flow-field characteristics of a large equivalent shallow-depth explosion is carried out using a self-developed Eulerian finite element solver. Firstly, the numerical model is validated against theoretical results and a small equivalent explosion test in a tank. The numerical results are found to agree well with the theoretical and experimental results. In the next step, the cavitation cut-off effect is added to the underwater explosion model, and the cavitation phenomenon is quantitatively analyzed through the flow-field pressure. In addition, the dynamic characteristics of the bubble and water hump under various initial conditions for different stand-off parameters are analyzed. The effect of gravity on these physical processes is also discussed. The bubble pulsation period, taking into account the free surface effect, is then quantitatively studied and compared with Cole's experimental formula for an underwater explosion. Overall, when the stand-off parameter $\gamma > 2$, the influence of the free surface on the empirical period of the bubble is not significant. Our investigation provides broad insights into shallow-depth underwater explosions from theoretical, experimental, and numerical perspectives.

## 1. Introduction

Bubble dynamics[1-6] are of enormous practical importance in multi-phase fluid mechanics and have a wide range of applications in many fields, such as underwater explosions,[7-10] the exploration of ocean resources,[11, 12] ultrasonic cleaning,[13, 14] and medical treatment.[15, 16] The explosive bubbles generated by an underwater explosion can cause severe damage to ships, marine structures, and habitats.[17-18] The pulsed pressure wave emitted by an air-gun-generated bubble can be reflected by complex seabed topography.[11,12] Micro-bubbles can efficiently clean the surfaces of structures.[13, 14] Extracorporeal shock wave lithotripsy (ESWL) is an effective medical treatment used to provide relief from kidney stones. High-energy shock waves are used to generate cavitation bubbles and their collapse can fragment a kidney stone into pieces.[15, 16] The dynamic characteristics of bubbles generated by underwater explosions are more complex and have a high level of research significance, due to their ability to displace a significant water volume and generate extreme wave conditions near the explosion, thereby enhancing demolition conditions. A great deal of research has been carried out on the engineering of underwater explosions.[18, 19]

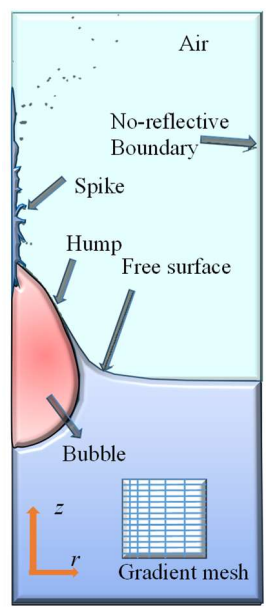

Fig. 1 A schematic diagram of a shallow-depth explosion.

The phenomenon of a bubble collapsing in a cuvette filled with water is called an underwater collapse of a bubble. Bubble characteristics are inherently complex, and their behavior becomes even more intricate, considering multiple bubbles and boundary effects[20, 21]. Wardlaw & Luton[22] and Rapet et al.[23] observed jets and shock waves reflecting off bubbles during the implosion of an underwater cavity near a structure. Zhang et al.[9] originally proposed an fully coupled fluid structure interaction model, and revealed the strong nonlinear interaction characteristics between underwater explosion bubble and structure. Han et al.[24] effectively combined BEM models and experiments, and uncovered the mechanism of the interaction between the cavitation bubbles and the interface of two immiscible fluids. Riley et al.[25] performed Eulerian simulations and Li et al.[26] conducted simulations based on the Navier–Stokes equations. Brett & Yiannakopolous[10] found that shock waves dominate the load on the steel plate, rather than pressure waves caused by pulsating bubbles. The loads induced by collapsing bubbles on a cylinder were more severe than other loadings.[23]

A shallow-depth explosion is a more complex multi-phase flow problem that has attracted the attention of a large number of researchers[27-29] because of its complexity, volatile flow, and effectiveness in destroying targets. Over the past century, a lot of excellent work has been done in the field of underwater explosions. However, there are still some technical difficulties, such as those posed by multi-scale simulations. The explosive shock wave, pulsating shock wave, and bubble jet load can all affect the vitality of marine ships. The bubbles and water hump are two typical flow characteristics observed during shallow-depth explosions. In addition, different flow dynamics characteristics are produced by different initial conditions. The water hump also has essential research value due to its destructive applications. It is formed specifically from shallow-depth underwater explosions by the pulsation of a bubble and the pressure gradient over the bubble's surface. A bubble's hump dynamics can provide a low-cost, universal terminal defense system for ships and marine structures. In essence, it is beneficial to master the key elements of shallow-depth explosion dynamics. A schematic diagram of a shallow-depth explosion is shown in Fig. 1.

The complexity of the shallow-depth explosion problem makes accurate analytical solutions challenging to obtain. Numerical simulations[26] and experimental studies[30] can help us to understand the physical essence of complicated problems. The traditional BEM[31, 32] struggles to represent refined phenomena, such as detailed fluid fragmentation. Other numerical methods have been used in the study of shallow-depth explosion problems, including the finite volume method (FVM),[26] the finite element method (FEM),[33] and the meshless method.[34] Szymczak et al.[28] established the shallow-depth explosion model, which does not consider the compressibility of the fluid. However, when compared to experimental results, their numerical model was found to be capable of roughly simulating the entire underwater explosion process. Li et al.[35] used *MSC Dytran* software to simulate bubble and free surface dynamics in shallow underwater explosions, proving that a Eulerian solver can deal with nonlinear problems. Li et al.[26] used the open-source solver *Open-FOAM* based on the FVM to simulate underwater explosions. The numerical results of the FVM are highly consistent with experimental results. Phan et al.[36] conducted a meticulous study on this problem using the FVM, demonstrated the complex hydrodynamic characteristics of the free surface, and determined the typical physics behind the unstable water skirt phenomenon for the first time in the academic field.

Furthermore, the interaction between bubbles and the free surface has also been studied extensively in experiments. Cole et al.[37] provided observations and empirical formulas for large underwater explosions using a number of experiments. Cui et al.[18] used a pressure-reducing device to analyze bubble characteristics under large buoyancy parameters. Zhang et al.[38] investigated the motion evolution processes of bubbles and water hump characteristics for different stand-off distances through small-scale electric spark bubble experiments. Despite all of this excellent previous work, some complicated problems still need to be solved, such as the characteristics of large equivalent shallow-depth explosions, the cavitation effect, and refined numerical simulations.

In this work, using state-of-the-art approaches, a transient numerical shallow-depth explosion model is established based on the compressible Eulerian FEM. Numerical results are obtained from the in-house codes developed by our team. Firstly, the numerical model proposed in this paper is verified through comparisons with theoretical predictions and with experimental results of a small equivalent explosion test in a meter-scale tank. The numerical results agree well with both the theoretical and experimental results. In the next step, the underwater explosion model, which takes into account the cavitation cut-off effect, is quantitatively analyzed by investigating the flow-field pressure. The dynamics of the bubble and water hump under three initial conditions for different stand-off parameters are then analyzed and presented. Finally, the effect of gravity on this physical process is further investigated. In addition, the bubble pulsation period considering the free surface effect is quantitatively compared with Cole's underwater explosion empirical formula.

## 2. Numerical methods

### 2.1 Eulerian finite element method

Numerical methods based on the Eulerian and Lagrangian perspectives are two typical approaches in computational fluid dynamics. Considering the advantages of both, the Eulerian finite element method[39-43] is an effective solution for dealing with complicated transient non-linear problems. A schematic diagram of this method is shown in Fig. 2. The fluid properties are assigned to the mesh in the Eulerian finite element method. The core of the strategy is to perform a Lagrangian calculation step, followed by a remapping in which the solution is converted from distorted Lagrangian information to Eulerian information.[44]

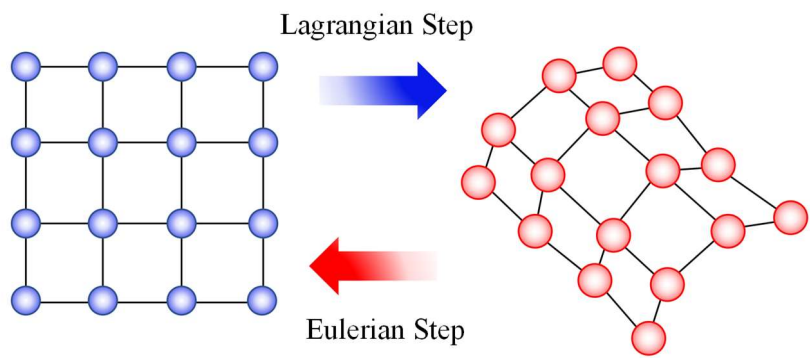

Fig. 2 A schematic diagram of the Eulerian finite element method.

The compressibility of the fluids is taken into account in the present work. However, the viscosity and surface tension of the fluid are ignored, as they have little effect on the overall transient shock process. The governing equations are the following — the continuity equation, the momentum equation, and the energy equation, respectively:

$$\begin{cases} \dfrac{\partial \rho}{\partial t} + \nabla \cdot (\rho \mathbf{u}) = 0 \\ \dfrac{\partial \rho \mathbf{u}}{\partial t} + \nabla \cdot (\rho \mathbf{u}\mathbf{u}) = -\nabla p + \rho \mathbf{b} \\ \dfrac{\partial \rho e_{in}}{\partial t} + \nabla \cdot (\rho e_{in} \mathbf{u}) = -p \nabla \cdot \mathbf{u} \end{cases} \qquad (1)$$

where $\rho$ denotes the fluid density, $t$ denotes the time, $\mathbf{u}$ denotes the velocity vector, $p$ denotes the element pressure, $\mathbf{b}$ denotes the body force, and $e_{in}$ denotes the internal energy per unit mass.

Operator splitting is the important mathematics thoughts, and the solution of the governing equation can be divided into Lagrangian and Eulerian calculation steps based on this method. The entire process is summarized as follows:

(1) The computational mesh is embedded in the material domain in the Lagrangian calculation step and the mesh position at discrete points is solved in time.

(2) When the pressure and gravity in each zone or element are known, the nodal forces are calculated. Immediately, the nodal acceleration is obtained by dividing the nodal forces by the nodal masses.

(3) In the Lagrangian calculation step, the velocity and displacement are given by integration.

(4) The volume of fluid (VOF) method, the monotonic upwind scheme for the conservation laws, and the half-index shift algorithm are used to determine the amount of transport between adjacent elements and nodes.

(5) Based on the work done, the internal energy, pressure, and density are recalculated.

(6) Variable correction processing is performed, such as for pressure balance, and the element variables are calculated again. Then, the new parameters are substituted into step (2), and the iterative calculation is continued until the calculation termination condition is met.

## 2.2 The equation of state

Table 1: Specific material constants for water and air.[45]

| Material | $P_w$ MPa | $\kappa$ | $\rho_0$ Kg/m$^3$ |
|---|---|---|---|
| Water | 330.9 | 7.15 | 1000 |
| Air | 0 | 1.25 | 1.40 |

Table 2: Specific material constants for TNT production.[45]

| Material | A GPa | B GPa | $R_1$ | $R_2$ | w | $\rho_0$ Kg/m$^3$ |
|---|---|---|---|---|---|---|
| TNT Production | 371.2 | 3.20 | 4.15 | 0.95 | 0.30 | 1630 |

The equation of state for the water and air is set from the Tammann equation:[12]

$$p = \rho e_m (\kappa - 1) - \kappa P_w, \tag{2}$$

where $P_w$ is the reference pressure. The pressure of the internal gas production of the explosive is modeled with the Jones–Wilkens–Lee (JWL) equation:[46]

$$p = A(1 - \frac{w\rho}{R_1 \rho_0}) e_1 + B(1 - \frac{w\rho}{R_2 \rho_0}) e_2 + w\rho e_m, \tag{3}$$

where $A$, $B$, $R_1$, $R_2$, and $w$ are the material constants. $\rho_0$ and $\rho$ are the densities of the explosive and the gas production. $e_1$ and $e_2$ are defined as

$$e_1 = \exp(-\frac{\rho_0}{\rho} R_1) \tag{4}$$

$$e_2 = \exp(-\frac{\rho_0}{\rho} R_2). \tag{5}$$

The fluid constants for the Tammann equation and the JWL equation are given in Table 1 and 2.

Upon the detonation of the explosive, a spherical shock wave moves outward. The free surface reflects this shock wave as a rarefaction wave, and it then travels back down through the gas globe of the explosive product. Due to the tension created behind the rarefaction wave, cavitation occurs, and some of the water is gasified. In science, cavitation is a relatively complex physical phenomenon. Various methods[47, 48] have been used to establish a reasonable cavitation model. The pressure truncation model and the cavitation model considering phase transitions[49] are two mainstream numerical models. The pressure truncation numerical model proposed by Xie et al.[48] is used in this paper:

$$P = \begin{cases} \text{Tammann's}, & P \geq P_{sat}, \\ P_{sat} + P_{gl} \cdot In\left[\dfrac{\rho_g c_g^2(\rho_l + \alpha(\rho_g - \rho_l))}{\rho_l(\rho_g c_g^2 - \alpha(\rho_g c_g^2 - \rho_l c_l^2))}\right], & P_\varepsilon < P \leq P_{sat}, \\ P_\varepsilon, & P < P_\varepsilon \end{cases} \quad (6)$$

where $P_{sat}$ represents the physical saturated vapor pressure, $P_\varepsilon$ represents a pressure close to zero, $P_{gl}$ represents the pressure phase associated with the volume fraction (with the subscript representing the gas and liquid phases), $\alpha$ represents the volume fraction of the liquid phase, and $c$ represents the velocity of sound. The numerical simulation results take into account the cavitation effect and will be presented in the following sections.

### 2.3 The multi-phase VOF method

The VOF method[50-53] can explicitly capture interfaces and describes the parameter of materials in terms of their volume fractions. It is one of the most common methods used in the area of computational fluid dynamics. For the physical problem studied in this paper, there are three fluids — air, water, and explosive production. When there is only one material in the element, the fluid interface is the boundary of the element. When there are two or three materials in the element, the fluid interface is determined by the least-squares method. In the current work, the air, water, and explosive production are presented as one element, and the volume fractions of the air and the explosive production are added together to produce a mixed material. This element is then treated as a two-phase element, as shown in the schematic diagram of the interface establishment (see Fig. 3). The governing equation for the volume fraction is:

$$\frac{\partial f_i}{\partial t}+u\cdot\nabla f_i=0, \tag{7}$$

where $f_i$ is the volume fraction of the fluid. The least-squares method equation is as follows and is used to determine the fluid interface:

$$\frac{\partial}{\partial m_i}\sum_j[x(\eta_j,\varsigma_j)-f_{air+products}]=0, \tag{8}$$

where $m_i$ is the coefficient of the interface function and is used to establish the interface, and $\eta_j$ and $\varsigma_j$ are the $j^{th}$ adjacent element center coordinates.

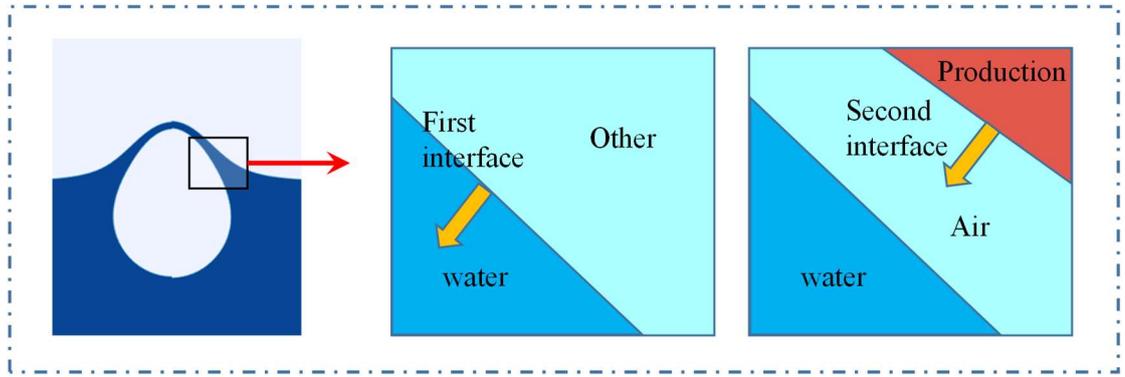

Fig. 3 A schematic diagram of the interface establishment for various fluid phases.

## 2.4 Numerical setup and parameter processing

All the current calculations are based on our team's self-developed in-house Eulerian finite element solver. The source code used for the calculation is written in the programming language FORTRAN, which is an assembly language that is widely used in scientific computing. In the numerical model, the computational cost is effectively saved by a parallel algorithm. In this work, the number of grid elements for each numerical case varies between 1 and 3 million, and this grid resolution was sufficient to achieve the reliable numerical simulation of the fluid dynamics characteristics. The simulations were performed using a processor Core i7-8700 (3.2 GHz) and a memory of 15.8 GB. The parallel Open-MP mode was used in our computational code to accelerate our calculations. The numerical simulation took between 5 and 24 h to compute the dynamics of an underwater explosive bubble up to more than two periods. In addition, an asymptotic grid and a non-reflective boundary were used in this work to further improve the simulation.

For a real-scale shallow-depth explosion, $W$ is the initial explosive weight, $d$ is the detonation water depth, and $R_m$ is the maximum diameter of the bubble. These are the dimensional parameters used to quantitatively describe the initial conditions. The non-dimensionless parameter $\gamma = d/R_m$ is used to qualitatively summarize the physical behaviors of the bubble and illustrate the results.

## 3 Comparison of the numerical results

### 3.1 Comparison with theory

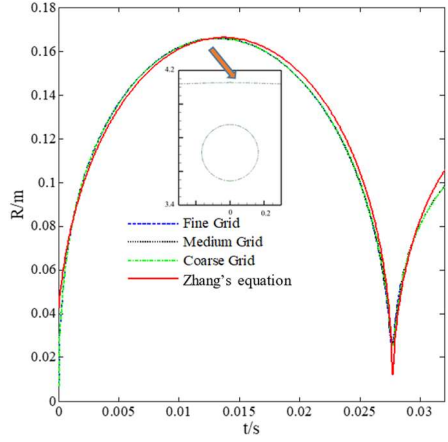

Fig. 4 A comparison of the numerical and theoretical results (Zhang's Equation[5]), the maximum bubble shapes with different grid sizes are marked.

Firstly, the numerical method was validated using the Zhang equation, which is an epoch-making milestone work in the field of bubble theory after the Rayleigh-Plesset and Keller equations, that can pioneeringly predict the compressible migration motion of bubbles.[5] When the bubbles interact with the free surface, they undergo a clear compressible migration, and Zhang's equation can express these complex characteristics. The equation for the bubbles is as follows, where $H_u$ is the enthalpy, $R$ is the bubble radius, $C$ is the sound velocity in water, and $v$ is the bubble velocity:

$$\left(\frac{C-\dot{R}}{R} + \frac{d}{dt}\right)\left[\frac{R^2}{C}\left(\frac{1}{2}\dot{R}^2 + \frac{1}{4}v^2 + H_u\right)\right] = 2R\dot{R}^2 + R^2\ddot{R} \qquad (9)$$

For comparison purposes, 1.3 g of TNT was detonated at a water depth of 0.4 m, generating bubbles with a maximum radius of 0.169 m. A comparison between the numerical and theoretical results is shown in Fig. 4. In the numerical simulation, the calculation domain was 0.5 m×1.0 m, and the grid sizes were 0.009$R_m$, 0.012$R_m$, and $0.015R_m$. The calculation boundary was set as a non-reflective boundary.[54] The accuracy and convergence of the numerical methods can be preliminarily verified through theory, and the bubble shape is also convergent. But for important phenomena such as free surface breaking and water humps, further verification is needed through experiments.

**3.2 Comparison with experiments**

After the comparison with theoretical predictions, the results of our numerical method were verified through a small equivalent explosion experiment. The small equivalent underwater explosion is an alternative experiment that is similar to a real-scale underwater explosion. Due to its smaller size, the associated costs are significantly reduced, and in addition, these experiments can provide improved personal and environmental safety. The free surface evolution and bubble pulsation behaviors can be obtained from the experiments by means of high-velocity videography. The experiments were carried out in a dedicated tank for explosive experiments. The size of the tank was 4 m×4 m×4 m, with a 10 mm wall thickness made of steel plates welded together to form the appearance of a cubic tank (see Fig. 5). And the water tank can only perform some small equivalent experiments, approximately within 50g of TNT explosive, in order to avoid the influence of boundaries. The front of the tank had an observation window and the illumination window on the opposite wall was made of transparent PMMA plates. A studio with a high-velocity camera and processing unit was set up on the side of the observation window. The illumination was provided by a 300W studio light to facilitate the imaging. The safe house served as space for scientists to be kept safe from the resulting shock wave, noise, and possible accidents. Due to the large-scale height, the studio and tank had a ladder. The fluid shape was captured through the high-velocity camera in the studio. A diagram of the experimental device is shown in Fig. 5.

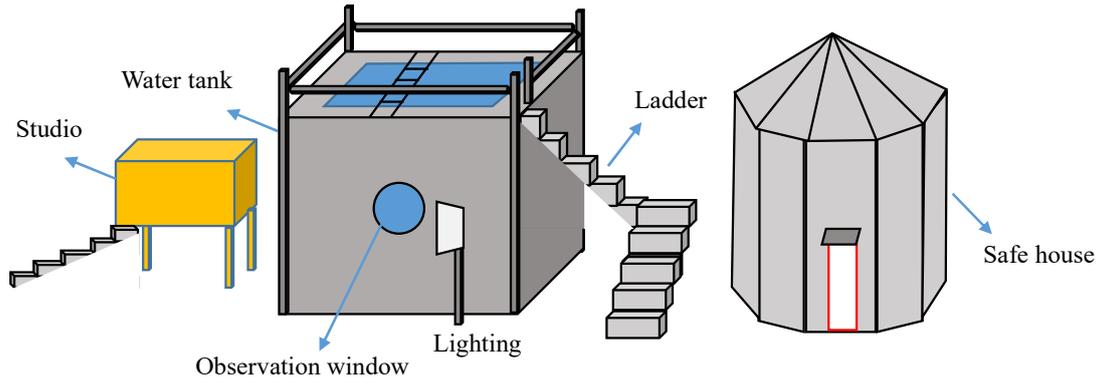

Fig. 5 A diagram of the experimental device.

Figure 6 shows a comparison of the experimental and numerical results for the free surface under typical shallow-depth underwater explosion conditions, in which 10.4 g of a TNT-equivalent explosive was detonated 3 cm beneath the free surface. In the numerical simulation, the computational domain had a size of 1.0 m×2.0 m and the number of grid elements was 125750. The outlet side boundaries were set as non-reflective boundaries.[54] It can be seen from Fig. 6 that the numerical results in this paper are in good agreement with the experimental results, showing similar typical physical characteristics for the free surface movement, bubble pulsation, and water hump. The actual process of the explosion is more complicated, so the free surface in water tank of the repeated experiments could not be completely calm. Therefore, the irregular asymmetric movement of the free surface in the later period was not simulated.

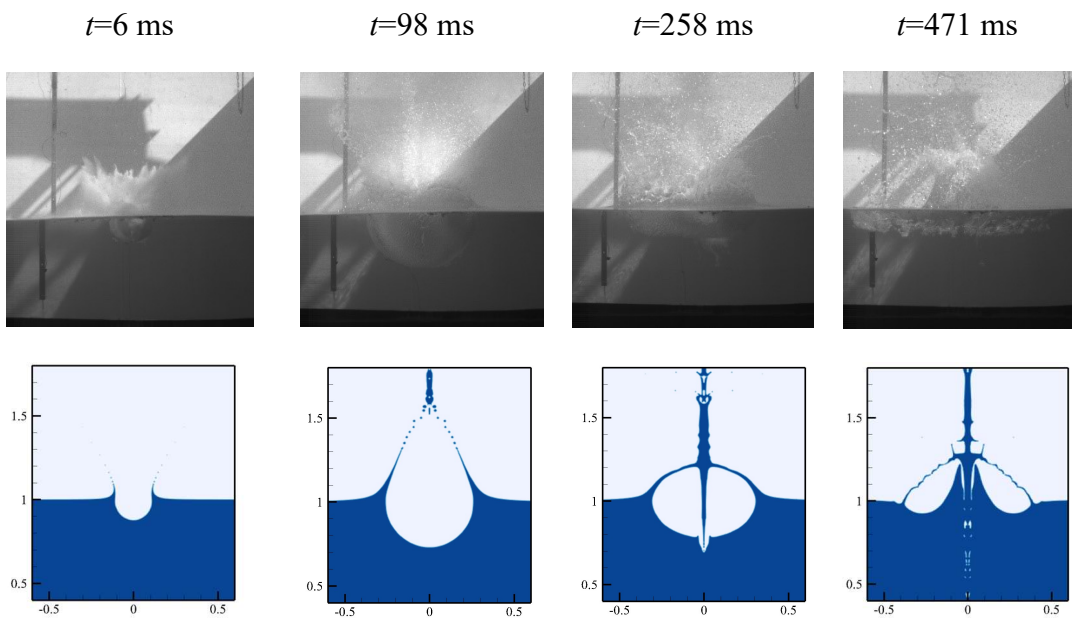

Fig. 6 A comparison of the experimental and numerical results for the free surface

under typical shallow-depth underwater explosion conditions in which 14.0 g of a TNT-equivalent explosive was detonated 2 cm beneath the free surface.

Overall, the results obtained from our proposed numerical method agree well with both the theoretical and experimental results. In the next section, the results of a few more numerical cases are illustrated under full-scale conditions.

## 4 Results and discussion

This section describes the results obtained from the proposed numerical method and analyzes the effect of cavitation models, the influence of the stand-off distance, the free surface, and gravity at full scale.

### 4.1 The influence of the cavitation model on the numerical simulation

Figure 7 shows a pressure and cavitation area diagram for 200 kg of explosive detonated 4 m beneath the free surface. As can be seen from Fig. 7, the cavitation phenomenon is clearly seen in the numerical simulation results. The cavitation area is marked in the figure as the area where the pressure is smaller than $P_\varepsilon$. For the large-equivalent shallow-depth underwater explosion problem, the existence of the cavitation phenomenon is undeniable. In contrast with previous studies, it is observed that the cavitation region evolves with the spread of the shock waves in the domain. Just after the generation of the rarefaction wave, the cavitation area is located above and close to the exploding bubble. Thereafter, due to the radiation of the wave generating a low-pressure zone, the cavitation zone is mainly distributed on both sides of the exploding bubble. Transmission is also clearly seen.

The time evolution of the pressure at three measurement points is shown in Fig. 8. The three measurement points, numbered 1 to 3, are 20 m away from the free surface and at horizontal distances of 7.5 m, 15 m, and 22.5 m from the bubble center position, respectively. As can be seen from Fig. 8, a large shock wave is detected at the measuring point P1. The peak value of the pressure differs greatly for different measuring points because the shock wave dissipates faster in the water. Subsequently, there is a distinct low-pressure zone in the curve caused by pressure truncation. At the end of the curve, cavitation collapse and rebounding occur, which is consistent with what has been described in previous literature[55] and fully validates the algorithm's effectiveness. The highest pressures measured during the explosion were about 12 MPa. Peaks in the pressure measurements were also observed post-explosion at $t$=0.82 s. These post-explosion pressure pulses can be related to the cavitation phenomenon because the cavitation regime also appears at the same time. And the cavitation phenomenon mostly occurs in the shock wave stage and has a short duration. This paper believes that its impact on the pulsation of explosives bubbles can be ignored.

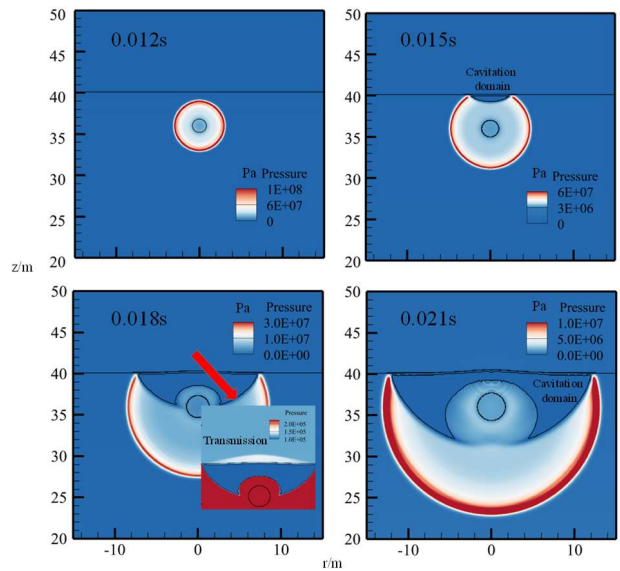

Fig. 7 A pressure and cavitation area diagram for 200 kg of explosive detonated 4 m beneath the free surface.

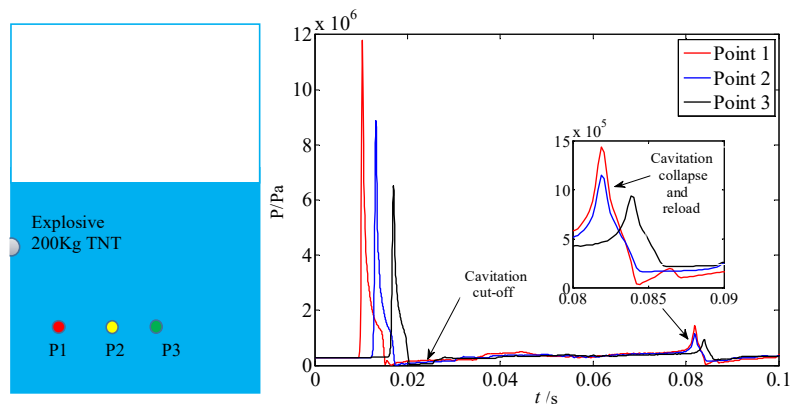

Fig. 8 The time evolution of the pressure measured at different measuring points in the domain.

### 4.2 The influence of the stand-off distance on the shallow-depth explosion

The stand-off distance is an important factor in shallow-depth underwater explosions and affects the flow field's dynamic characteristics. For real-scale underwater explosion problems, the bubble pulsation characteristics and the water hump shapes are more complicated than those of small bubbles. Meanwhile, the effect of the stand-off distance has a more significant effect on the explosion and surrounding fluid-flow characteristics. The numerical simulation results under three typical initial conditions are analyzed in this article and the corresponding explosion physics is summarized. In order to thoroughly understand the flow field dynamics characteristics as a whole, fixing of the upper and lower limits is adopted in our work for velocity diagrams with the same working conditions. And the size of the computational domain is all $50m \times 100$m, with a minimum grid size of 0.1 m and a non-reflective boundary.

When $\gamma = 0.11$, the initial position of the explosive is very close to the free surface, The free surface is ruptured and reclosed during the expansion and collapse stage of the bubble. Tian et al. provided a detailed introduction to this phenomenon for near-surface exploding bubbles,[33] and a velocity diagram shown in this paper further reveals the underlying physical mechanism. As shown in Fig. 9, the breaking and reclosing of the free surface is clearly reflected in the computed results. It can be postulated that the high-pressure gas generated by the instantaneous explosion diffuses to the surroundings at a very high velocity, breaking the free surface. When the free surface recloses to form a new bubble, the intense motions of the free surface still produce an outward-moving jet. When $t = 0.86$ s, the bubble enters the annular bubble stage due to the jet penetration. Owing to the internal-external pressure difference and buoyancy, the thin water layer breaks in the second expansion stage of the bubble at $t = 1.25$ s. In addition, the evolution of a water hump under such conditions is clearly seen after $t = 1.97$ s. At this time, the water hump is relatively thin, and its height is exceptionally high, reaching a height of the order of hundreds of meters. In the bubble expansion stage, since the bubble is close to the free surface, the rate at which explosion energy is obtained by the water at the free surface is very significant. Due to the breaking of the water layer, a water skirt is also formed at the root of the water hump. This phenomenon has rarely been addressed in previous investigations in the literature. At $t = 1.97$ s, the higher precision of the numerical calculation method enables the feathering and fragmentation phenomenon of the free surface to be effectively captured. This feature demonstrates the ability of the proposed numerical method to capture tiny features of the flow dynamics during underwater explosions.

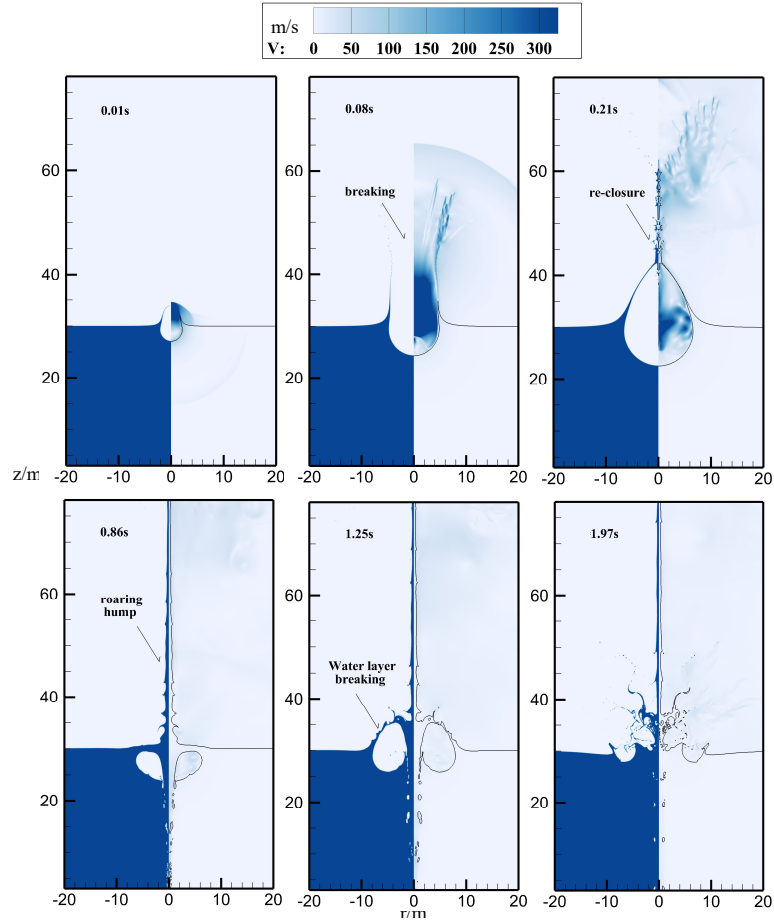

Fig. 9 The evolution of the free surface and velocity field for the case in which 200 kg of explosive is detonated 2 m beneath the free surface. The free surface scalar is shown in the left half and the velocity scalar is shown in the right half of each panel. Very high velocities are observed during the explosion stages, which break the free surface.

When $\gamma=2.22$, the initial position of the explosive is relatively far from the free surface compared to the previous case. There exists very little detailed analysis of this kind of underwater explosion in the previous literature. It can be seen from Fig. 10 that the bubble expands spherically in the expansion stage after explosion, and that the shock wave generated by the explosion is transmitted and reflected at the free surface. The transmission mode of the shock wave was described in detail in the above section.

Since the induced force generated by the free surface is smaller than the buoyancy of the bubble, in this case, the bubble generates an upward jet, and the dynamics of the bubble are initially consistent overall with those of a free-field explosion bubble. After $t=0.78$ s, the bubble enters the annular stage, and the bubble exhibits a floating motion state in the form of an annular bubble due to the effect of buoyancy. At $t=1.79$ s, the bubble breaks at the free surface, forming a water skirt feature. In general, this water hump is low in height, but its bottom shape is very similar to the water hump in Fig. 9. Since the bubble eventually produces a significant dynamic response to the free surface, this case can also be considered a typical shallow-depth explosion.

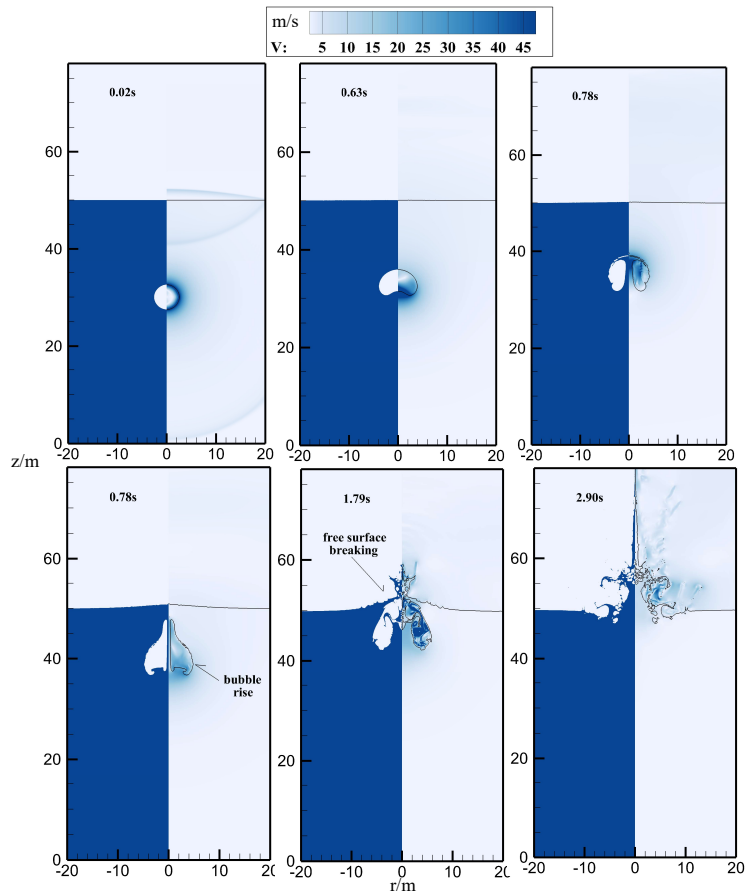

Fig. 10 The evolution of the free surface and velocity field for a case in which 200 kg of explosive is detonated 20 m beneath the free surface. A floating bubble and water hump are observed.

When $\gamma=0.28$, the initial position of the explosive is a little farther from the free surface. Under this condition, the expansion of the bubble causes the free surface to rise, as shown in Fig. 11. When the bubble grows to a definite volume, a jet similar to that in Fig. 9 is generated due to the induction of the free surface, and the generation mechanisms of the two are the same. At $t=0.90$ s, the bubble produces a rebound effect in the annular phase, and the rebound causes fluid in the free surface to obtain a certain velocity of movement, which in turn causes two tiny water skirts to appear on both sides of the water hump. As shown in Fig. 11, the water skirt movement causes the water hump to adopt a fascinating crown shape. When $t=1.70$ s, the number of water skirts on both sides continuously increases due to multiple pulsations of the bubbles. This phenomenon has also rarely been described in previous literature. In addition, when comparing the current results with the results of the first two cases with different initial conditions, it is seen that the width of this water hump is larger. For engineering applications, this form could be more suitable for the demolition of ships and anti-ship missiles.

The time evolution of the height of the water hump under different stand-off distances is shown in Fig. 12. The height $H$ of the water hump as defined in this paper is the height of the water point with the most significant vertical height from the free surface. It can be seen from the figure that the change in the height of the water hump varies significantly when $\gamma$ varies between 0.11 and 2.22. Smaller stand-off distances correspond to larger water hump heights, and the figure intuitively shows the evolution of the water hump.

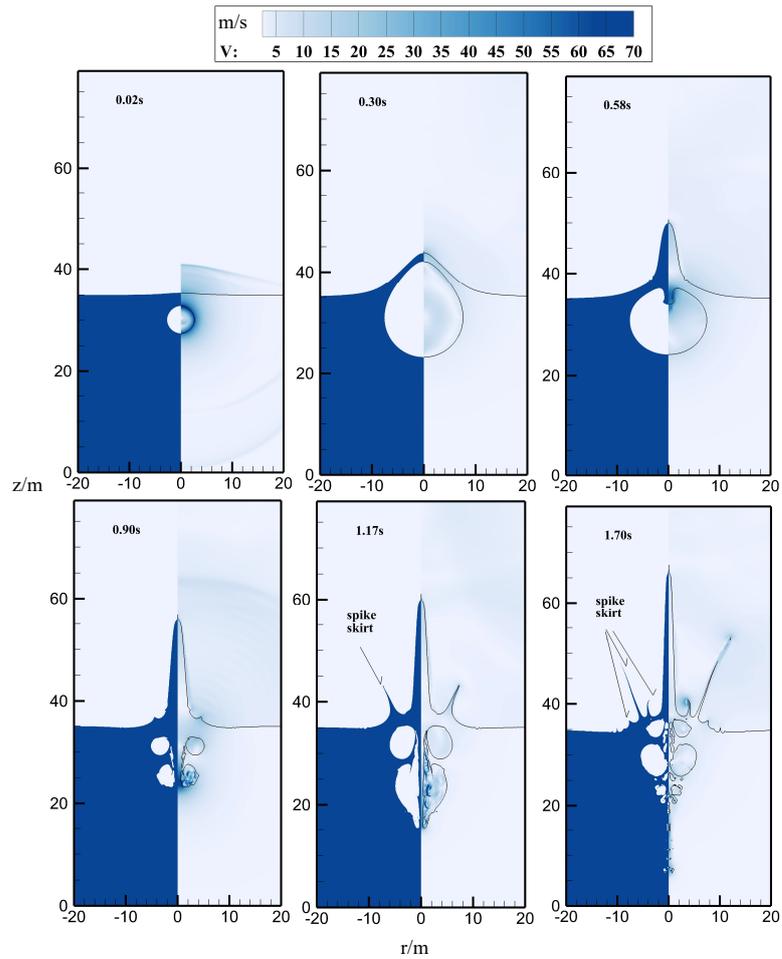

Fig.11 The evolution of the free surface and velocity field for the case in which 200 kg of explosive is detonated 5 m beneath the free surface.

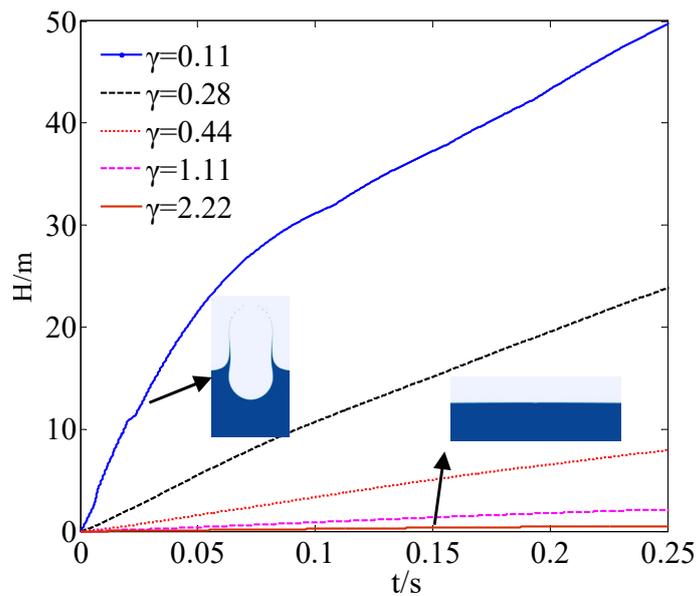

Fig. 12 The evolution of the height of the water hump under different stand-off parameters.

Meanwhile, the height of the water hump is reduced for larger stand-off distances. In our investigations, we found that $\gamma = 0.11$ and $\gamma = 0.28$ resulted in the most significant hump characteristics. When $\gamma > 1$, the water hump heights in initial stage are relatively not more significant.

**4.3 The influence of gravity on bubble characteristics**

For large equivalent underwater explosion problems, the complexity of the dynamics of the explosion is primarily driven by the effect of gravity. It is difficult to observe the details of fluid motion and explosion dynamics in real-scale experiments of underwater explosions. Meanwhile, small laboratory-scale decompression bubble experiments have certain practical limitations.[18] Therefore, detailed numerical investigations are advantageous over experiments for studying some realistic underwater explosion scenarios. Numerical simulations were performed for 200 kg of explosive detonated 4 m beneath the free surface, both with and without gravity. Figure 13 compares the free surface evolution in the two cases. The effect of gravity on a large equivalent underwater explosion is clearly shown in Fig. 13.

It can be seen from Fig. 13 that in the initial expansion stage of the bubble, the buoyancy effect does not significantly affect the explosion dynamics, and gravity has little effect on the bubble growth and free surface evolution. When the bubble expands to a certain volume (see at $t = 0.44$ s in Fig. 13), gravity starts to play a significant role in the shapes of the bubble and hump. Meanwhile, the bubble shrinks and the corresponding relative volume decreases. On the other hand, the relative height of the water hump also decreases to some extent. When the jet penetrates the bubble, the bubble enters the annular stage. Under buoyancy, the left bubble is closer to the free surface and the bubble shape is more complicated. In the further rebound stage of the bubble, the water skirt phenomenon first appears at the free surface on the left side. The height of the left water skirt is higher due to more energy from the pulsating bubble being transmitted to the surface fluid. In addition, under the effect of gravity, the induction effect of free surface to bubbles is not obvious. The right bubble undergoes significant movement away from the free surface.

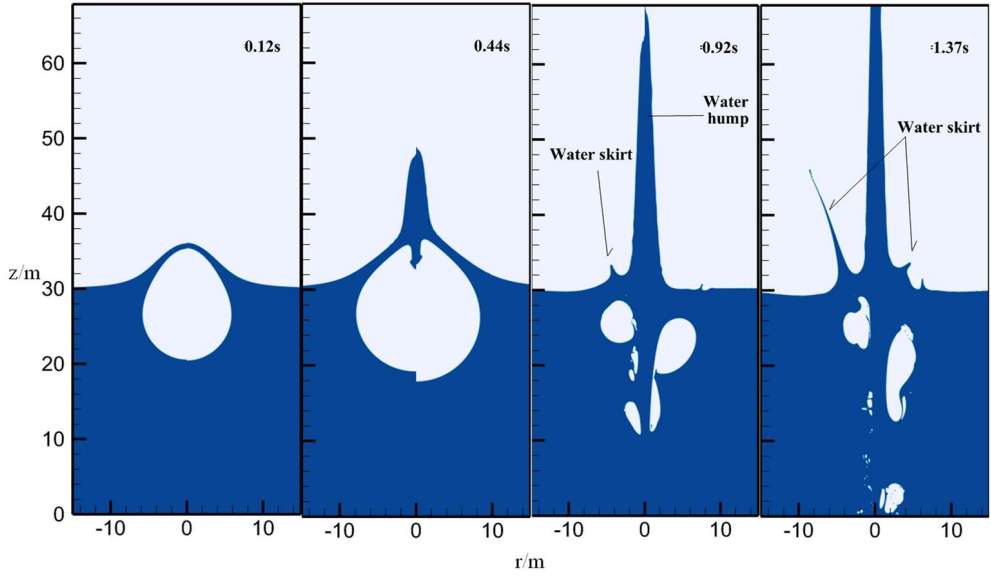

Fig. 13 A comparison of the free surface evolution with and without gravity for 200 kg of explosive detonated 4 m beneath the free surface.

The results of our investigation highlight that gravity must be taken into account when analyzing underwater explosions. Gravity significantly affects the free surface dynamics as well as the dynamics of the bubble's later collapse and rebound. In summary, a detailed analysis of the fundamental mechanisms behind underwater explosion problems has been carried out in this paper.

**4.4 The influence of the free surface on the bubble pulsation period**

The period of underwater bubble pulsation is a critical physical parameter. When the pulsation period of the bubble is consistent with the natural period of a nearby marine structure, such as a ship, the bubble and structure will resonate, and the structure may be severely damaged by the bubble's collapse dynamics. Through theoretical research and experimental research, an empirical formula was proposed by Cole[37] that can predict the pulsation period and the maximum diameter of the underwater explosion bubble:

$$T_p = 2.11 \frac{W^{\frac{1}{3}}}{(d+10.33)^{\frac{5}{6}}}, \qquad (10)$$

$$R_m = 3.36 \frac{W^{\frac{1}{3}}}{(d+10.33)^{\frac{1}{3}}}, \qquad (11)$$

where $T_p$ is the period of pulsation. The pulsation period changes significantly for a shallow-depth explosion bubble compared to the underwater explosion case for $\gamma < 2$, and can no longer be predicted by the empirical formula (see Fig. 14). Therefore, the influence of the free surface on the bubble pulsation is analyzed in this paper through numerical simulations.

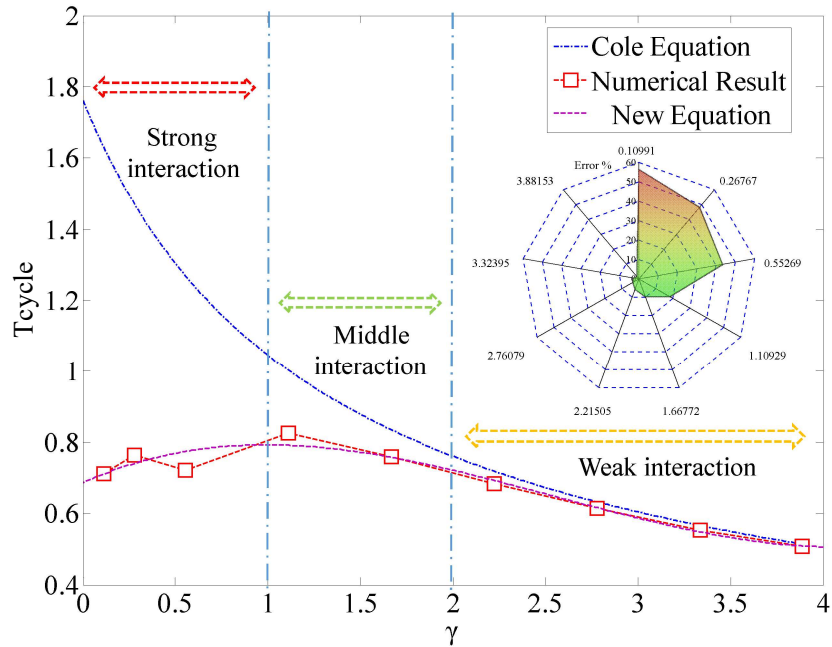

Fig. 14 A comparison of the pulsation period between the numerical simulations and the empirical formula proposed by Cole. A relative error diagram is shown in the sub-figure.

From Fig. 14, it can be concluded that when $\gamma > 2$, the influence of the free surface on the empirical period of the bubble is not significant. The estimated error of the empirical formula, in this case, is less than five percent. The closer the detonation is to the free surface, the larger the errors between the pulsation period predicted by the numerical simulation and Cole's empirical equation become. At this time, the dynamics of the bubble in the first oscillation period are similar to that of the bubble in the free field, and the bubble pulsation period can be forecast by using Cole's empirical formula. When $1 < \gamma < 2$, the bubble begins to be significantly affected by the Bjerknes forces of the free surface. Under the influence of free surface effects, the bubble period is shortened, and the evolution process of the bubble is accelerated. When $\gamma < 1$, Cole's empirical formula cannot predict the bubble motion state, resulting in errors of more than 30%. The change in the period exhibits a nonlinear trend. Meanwhile, the breaking and reclosing occur at the free surface in the bubble expansion stage. Therefore, the closer the initial position of the bubble is to the free surface, the smaller the maximum volume reached by the bubble is due to the dissipation of energy in the air, which affects the bubble. As the buoyancy effect is reduced, the Bjerknes force plays a more obvious role. Under the combined action of buoyancy and the Bjerknes force, the bubble gradually evolves. When the sum of the buoyancy and the Bjerknes force is greater, the bubble has a shorter oscillation period. Because of the strong nonlinearity of this process, the traditional linear equation cannot describe the physical features and is unable to predict reliable pulsating periods. In addition, when $\gamma < 1$, the shallow-depth explosion period formula is modified by combining it with the calculated data, given as:

$$T_P = 2.11 \frac{W^{\frac{1}{3}}}{(d+10.33)^{\frac{5}{6}}} (1-\beta(\gamma)), \tag{12}$$

$$\beta(\gamma) = a\gamma^3 + b\gamma^2 + c\gamma + d, \tag{13}$$

where $\beta(\gamma)$ is the correction function, which is considered to be a cubic function in this work. Using a curve fitting function, the coefficients can be obtained as *a*=0.86, *b*=-1.4, *c*=0.67, *d*=-0.65 Overall, our numerical results serve to compensate for the errors in Cole's formula and to predict the oscillation period for $\gamma > 2$. Our suggested modifications to Cole's empirical formula make it suitable for predicting the oscillation period of an underwater explosion bubble for a wide range of stand-off parameters covering shallow, middle, and deep underwater explosions.

**5. Conclusion**

A shallow-depth explosion transient numerical model based on the compressible Eulerian finite element method has been proposed in this paper. The pressure of the internal gaseous product of the explosive is modeled with the Jones–Wilkens–Lee (JWL) equation. The obtained results agree well with experiments and theory, proving that our proposed numerical model is reliable. The following conclusions can be drawn from our investigations and comparisons:

(1) The cavitation phenomenon is obvious in the large equivalent shallow-depth underwater explosion problem. The cavitation region during an underwater explosion evolves with the spread of shock waves in the explosion region.

(2) For $\gamma < 0.11$, the free surface breaks and recloses during the expansion stage of the bubble. Meanwhile, for $\gamma > 2.22$, the bubble moves as a spherical shape in the expansion stage. The width of the water hump is more significant and wider for $\gamma = 0.28$. We recommend this case for demolition or attacks on anti-ship missiles. In addition, the small stand-off parameter corresponds to a higher magnitude of the water hump height.

(3) In the initial expansion stage of the bubble, the buoyancy effect is not obvious, and gravity has little effect on the evolution of the bubble and the free surface. When the bubble expands to a certain volume, the water hump and bubble evolve in a complicated way due to the effect of gravity.

(4) For $\gamma > 2$, the influence of the free surface on the empirical period of the bubble is not obvious. The dynamics of the bubble in the first oscillation period are similar to that of a bubble in the free field, and the bubble motion period can be forecast by using Cole's empirical formula. Meanwhile, for $1<\gamma<2$, the bubble begins to be more significantly affected by the Bjerknes force of the free surface. For $\gamma<1$, the bubble oscillation period exhibits a nonlinear trend and the prediction of the oscillation period by Cole's formula is no longer valid. Therefore, our study proposes a correction strategy to Cole's empirical formula so that it can be used to predict the oscillation period of an underwater explosion bubble for a wide range of stand-off parameters.

Overall, our investigation has given a very broad insight into shallow-depth explosions and the effects of various parameters. It has also suggested a correction to Cole's empirical formula. For future work, it would be beneficial to carry out more work on experiments and 3D effects in numerical methods for both laboratory-scale and full-scale measurements.

**Acknowledgments**

This work was supported by funding from the Finance Science and Technology Project of Hainan Province (Grant No. ZDKJ2021020), the National Key R&D Program of China (Grant Nos. 2022YFC2803500), and the National Natural Science Foundation of China (Grant No. 51909041).

**Date Availability**

The data that support the finding of this study are available from the corresponding author upon reasonable request.

**References**

[1] S. Li, A.M. Zhang, R. Han, Q. Ma. 3D full coupling model for strong interaction between a pulsating bubble and a movable sphere. Journal of Computational Physics, 2019, 392: 713-731.

[2] P. Cui, A.M. Zhang, S.P. Wang, Y.L. Liu, Experimental study on interaction, shock wave emission and ice breaking of two collapsing bubbles, Journal of Fluid Mechanics, 897 (2020) 40.

[3] S.M. Li, A.M. Zhang, P. Cui, S. Li, Y.L. Liu. Vertically neutral collapse of a pulsating bubble at the corner of a free surface and a rigid wall. Journal of Fluid Mechanics, 2023, 962, A28.

[4] K. Murakami, R. Gaudron, E. Johnsen, Shape stability of a gas bubble in a soft solid, Ultrasonics Sonochemistry, 67 (2020) 12.

[5] A.M. Zhang, S.M. Li, P. Cui, S. Li, Y. L. Liu, A unified theory for bubble dynamics, Physics of Fluids, 35 (2023) 033323.

[6] A.M. Zhang, B.Y. Ni, Influences of different forces on the bubble entrainment into a stationary Gaussian vortex, SCIENCE CHINA Physics, Mechanics and Astronomy, 56 (2013) 8.

[7] P.H. Chavanis, B. Denet, M. Le Berre, Y. Pomeau, Implosion-explosion in supernovae, Epl, 129 (2020) 7.

[8] M. He, A.M. Zhang, Y.L. Liu, Prolonged simulation of near-free surface underwater explosion based

on Eulerian finite element method, Theoretical and Applied Mechanics Letters, 10 (2020) 16-22.

[9] A. M. Zhang, W. B. Wu, Y. L. Liu, Q. X. Wang; Nonlinear interaction between underwater explosion bubble and structure based on fully coupled model. Physics of Fluids, 2017; 29 (8): 082111

[10] J.M. Brett, G. Yiannakopolous, A study of explosive effects in close proximity to a submerged cylinder, International Journal of Impact Engineering, 35 (2008) 206-225.

[11] S. Zhang, S.P. Wang, A.M. Zhang, P. Cui, Numerical study on motion of the air-gun bubble based on boundary integral method, Ocean Engineering, 154 (2018) 70-80.

[12] H. Liu, K. Yang, Y. Ma, Q. Yang, C. Huang, Synchrosqueezing transform for geoacoustic inversion with air-gun source in the East China Sea, Applied Acoustics, 169 (2020).

[13] X. Huang, H.B. Hu, S. Li, A.M. Zhang, Nonlinear dynamics of a cavitation bubble pair near a rigid boundary in a standing ultrasonic wave field, Ultrasonics Sonochemistry, 64 (2020) 11.

[14] T. Tuziuti, Influence of sonication conditions on the efficiency of ultrasonic cleaning with flowing micrometer-sized air bubbles, Ultrasonics Sonochemistry, 29 (2016) 604-611.

[15] M. Postema, A. Bouakaz, Acoustic bubbles in therapy: Recent advances with medical microbubbles, clouds, and harmonic antibubbles Preface, Applied Acoustics, 140 (2018) 150-152.

[16] O.V. Muravieva, O.P. Bogdan, D.V. Zlobin, V.N. Milich, S.I. Maslennikov, Y.S. Dudina, Experimental evaluation of the ultrasound radiation intensity of medical equipment based on the analysis of the sizes of equilibrium gas bubbles in a liquid, Instruments and Experimental Techniques, 60 (2017) 137-142.

[17] E. Klaseboer, B.C. Khoo, K.C. Hung, Dynamics of an oscillating bubble near a floating structure, Journal of Fluids and Structures, 21 (2005) 395-412.

[18] A.M. Zhang, P. Cui, J. Cui, Q.X. Wang, Experimental study on bubble dynamics subject to buoyancy, Journal of Fluid Mechanics, 776 (2015) 137-160.

[19] A.M. Zhang, Y.L. Liu. Improved three-dimensional bubble dynamics model based on boundary element method. Journal of Computational Physics, 2015, 294: 208-223.

[20] N.N. Liu, A.M. Zhang, P. Cui, S. P. Wang, S. Li, Interaction of two out-of-phase underwater explosion bubbles, Physics of Fluids, 33 (2021) 106103.

[21] J. Yu, H.T. Li, Z.X. Sheng, Y. Hao, J.H. Liu, Numerical research on the cavitation effect induced by underwater multi-point explosion near free surface, AIP Advances, 13 (2023) 015021.

[22] A.B. Wardlaw Jr, J.A. Luton, Fluid-structure interaction mechanisms for close-in explosions, Shock and Vibration, 7 (2000) 265-275.

[23] J. Rapet, Y. Tagawa, C. D. Ohl, Shear-wave generation from cavitation in soft solids, Applied Physics Letters, 114 (2019) 123702.

[24] R. Han, A.M. Zhang, S. Tan, S.Li. Interaction of cavitation bubbles with the interface of two immiscible fluids on multiple time scales. Journal of Fluid Mechanics, 2022, 932, A8.

[25] M. Riley, M. Smith, J. Van Aanhold, N. Alin, Loading on a rigid target from close proximity underwater explosions, Shock and Vibration, 19 (2012) 555-571.

[26] T. Li, A.M. Zhang, S.P. Wang, S. Li, W.T. Liu, Bubble interactions and bursting behaviors near a free surface, Physics of Fluids, 31 (2019) 16.

[27] A.G. Soloway, P.H. Dahl, Peak sound pressure and sound exposure level from underwater explosions in shallow water, Journal of the Acoustical Society of America, 136 (2014) EL218-EL223.

[28] W.G. Szymczak, R.M. Gamache, Predictions of the Ejected Water Mass from Shallow Depth Explosion Plumes, International Journal of Nonlinear Sciences and Numerical Simulation, 17 (2016) 159-173.

[29] A. Daramizadeh, M.R. Ansari, Numerical simulation of underwater explosion near air-water free surface using a five-equation reduced model, Ocean Engineering, 110 (2015) 25-35.

[30] M. Greenhow, W.-M. Lin, Nonlinear-Free Surface Effects: Experiments and Theory, DOI (1983-09).

[31] A. Pearson, E. Cox, J.R. Blake, S.R. Otto, Bubble interactions near a free surface, Engineering Analysis with Boundary Elements, 28 (2004) 295-313.

[32] Z. Zong, J.X. Wang, L. Zhou, G.Y. Zhang, Fully nonlinear 3D interaction of bubble dynamics and a submerged or floating structure, Applied Ocean Research, 53 (2015) 236-249.

[33] Z.L. Tian, Y.L. Liu, A.M. Zhang, S.P. Wang, Analysis of breaking and re-closure of a bubble near a free surface based on the Eulerian finite element method, Computers & Fluids, 170 (2018) 41-52.

[34] P.N. Sun, D. Le Touze, G. Oger, A.M. Zhang, An accurate SPH Volume Adaptive Scheme for modeling strongly-compressible multiphase flows. Part 1: Numerical scheme and validations with basic 1D and 2D benchmarks, Journal of Computational Physics, 426 (2021).

[35] J. Li, J.L. Rong, Bubble and free surface dynamics in shallow underwater explosion, Ocean Engineering, 38 (2011) 1861-1868.

[36] P. Thanh Hoang, N. Van Tu, W. G. Park, Numerical study on strong nonlinear interactions between spark-generated underwater explosion bubbles and a free surface, International Journal of Heat and Mass Transfer, 163 (2020).

[37] R.H. Cole, R. Weller, Underwater Explosions, Physics Today, 1 (1948) 35-35.

[38] S. Zhang, A.M. Zhang, S.P. Wang, J. Cui, Dynamic characteristics of large scale spark bubbles close to different boundaries, Physics of Fluids, 29 (2017) 10.

[39] D.J. Benson, Momentum advection on unstructured staggered quadrilateral meshes, International Journal for Numerical Methods in Engineering, 75 (2008) 1549-1580.

[40] D.J. Benson, Momentum advection on a staggered mesh, Journal of Computational Physics, 100 (1992) 143-162.

[41] D.J. Benson, S. Okazawa, Contact in a multi-material Eulerian finite element formulation, Computer Methods in Applied Mechanics and Engineering, 193 (2004) 4277-4298.

[42] Y.L. Liu, A.M. Zhang, Z.L. Tian, S.P. Wang, Investigation of free-field underwater explosion with Eulerian finite element method, Ocean Engineering, 166 (2018) 182-190.

[43] Y.L. Liu, A.M. Zhang, Z.L. Tian, S.P. Wang, Dynamical behavior of an oscillating bubble initially between two liquids, Physics of Fluids, 31 (2019) 13.

[44] W. T. Liu, A.M. Zhang, X. H. Miao, F. R. Ming, Y. L. Liu, Investigation of hydrodynamics of water impact and tail slamming of high-speed water entry with a novel immersed boundary method, Journal of Fluid Mechanics, 958 (2023) A42.

[45] J. Qiu, T. Liu, B.C. Khoo, Simulations of compressible two-medium flow by Runge-Kutta discontinuous Galerkin methods with ghost fluid method, Communications in Computational Physics, 3 (2008) 479-504.

[46] E.L. Lee, H.C. Hornig, J.W. Kury, Adiabatic Expansion of High Explosive Detonation Products, DOI 10.2172/4783904(1968).

[47] J. Zhu, T. Liu, J. Qiu, B.C. Khoo, RKDG methods with WENO limiters for unsteady cavitating flow, Computers & Fluids, 57 (2012) 52-65.

[48] W.F. Xie, T.G. Liu, B.C. Khoo, Application of a one-fluid model for large scale homogeneous unsteady cavitation: The modified Schmidt model, Computers & Fluids, 35 (2006) 1177-1192.

[49] E. Stavropoulos-Vasilakis, N. Kyriazis, H. Jadidbonab, P. Koukouvinis, M. Gavaises, Review of

Numerical Methodologies for Modeling Cavitation, Cavitation and Bubble Dynamics, DOI (2021) 1-35.

[50] A. Albadawi, D.B. Donoghue, A.J. Robinson, D.B. Murray, Y.M.C. Delauré, On the assessment of a VOF based compressive interface capturing scheme for the analysis of bubble impact on and bounce from a flat horizontal surface, International Journal of Multiphase Flow, 65 (2014) 82-97.

[51] F. Denner, B.G.M. van Wachem, Compressive VOF method with skewness correction to capture sharp interfaces on arbitrary meshes, Journal of Computational Physics, 279 (2014) 127-144.

[52] C.W. Hirt, B.D. Nichols, Volume of fluid (VOF) method for the dynamics of free boundaries, Journal of Computational Physics, 39 (1981) 201-225.

[53] T. Li, S. Wang, S. Li, A.M. Zhang, Numerical investigation of an underwater explosion bubble based on FVM and VOF, Applied Ocean Research, 74 (2018) 49-58.

[54] C.A. Felippa, A family of early-time approximations for fluid-structure interaction, Journal of Applied Mechanics, Transactions ASME, 47 (1980) 703-708.

[55] W.B. Wu, A.M. Zhang, Y.L. Liu, S.P. Wang, Local discontinuous Galerkin method for far-field underwater explosion shock wave and cavitation, Applied Ocean Research, 87 (2019) 102-110.